\documentclass[conference]{IEEEtran}
\usepackage[latin9]{inputenc}
\usepackage{amsmath}
\usepackage{amssymb}
\usepackage{graphicx}
\usepackage{setspace}
\makeatletter
\usepackage{cite}
\usepackage{amsfonts}\usepackage{algorithmic}
\usepackage{textcomp}
\usepackage{xcolor}
\def\BibTeX{{\rm B\kern-.05em{\sc i\kern-.025em b}\kern-.08em
    T\kern-.1667em\lower.7ex\hbox{E}\kern-.125emX}}

\usepackage{setspace}
\makeatother

\begin{document}
\title{A N-Path Receiver With Harmonic Response Suppression\\
}
\author{Venkata Suresh Rayudu, Ki Yong Kim, and Ranjit Gharpurey\\
The University of Texas at Austin}
\maketitle
\begin{abstract}
A downconversion receiver employing a switch-based N-path filter with
reduced harmonic response around the third- and fifth- LO harmonics
is presented. The N-path filter is placed in a frequency-translation
feedback loop that is effective at the $3^{\text{rd}}$ and the $5^{\text{\text{th}}}$
LO harmonics to mitigate harmonic downconversion. A pulse-width-modulated
LO (PWM-LO) clocking scheme is used in the feedback upconverter to
reduce the noise injected around the LO harmonic at the input of N-path
downconverter. The compression resulting from blockers around the
$3^{\text{rd}}$ and the $5^{\text{\text{th}}}$ LO harmonics is also
suppressed as a result of reduced harmonic response. Compensation
of peak frequency shift of the N-path response due to parasitic input
capacitance is also described. \bstctlcite{IEEEexample:BSTcontrol}
\end{abstract}


\section{Introduction}

A switched-capacitor N-path circuit\cite{Franks1960} can be employed
for filtering an RF signal as well as a passive downconverter. Switch-based
implementations of such filters have been demonstrated in recent literature,
e.g., \cite{Cook2006}. N-path filters can provide a high effective
quality factor, a high dynamic range and a tunable response, in addition
to being scalable with process technology \cite{Darvishi2013}.

A known limitation of a switch-based N-path filter is that in addition
to downconverting signals around the desired center frequency, the
circuit also downconverts signals located around harmonics of the
center frequency. An approach to reduce the harmonic response employing
a PWM representation of a sinusoidal clock was shown in \cite{RayuduPWMNPATH}.
An N-path filter with a fundamental response at $f_{LO}$ that employs
rectangular clocks requires a primary clock frequency of $Nf_{LO}$.
With PWM clocks, the primary clock frequency is $MNf_{LO}$, where
the degree of harmonic suppression improves with increasing $M$.
For $N=4$ and $M=4$, the approach effectively removes harmonic response
up to nearly \textit{$8\times f_{\text{\textit{LO}}}$ }as shown in
\cite{RayuduPWMNPATH}.\textit{ }A challenge of the approach is the
requirement for high PWM clock frequency, which can limit the achievable
harmonic suppression due to narrow pulse-widths.

Mixer-first receivers that provide cancellation of LO harmonic responses
at the receiver output have been demonstrated in prior literature
\cite{andrews_passive_2010,murphy_noise-cancelling_2015,xu_blocker-tolerant_2018}.
The goal of this work is to provide harmonic response suppression
at the input of the N-path filter, similar to \cite{RayuduPWMNPATH},
without the requirement for high clock frequency. A frequency-translation
feedback loop is used in the proposed approach. The use of a frequency-translation
loop for input matching and enhancing out-of-band blocker performance
near the fundamental LO frequency of a receiver was shown in \cite{he_compact_2011}.
Feedback-based interference-rejection in a broadband channelizer using
reconfigurable harmonic rejection mixers in the feedforward and feedback
paths for suppression of harmonic LO levels was described in \cite{ho_active_2013}.

\begin{figure*}[t]
\centering{\includegraphics[height=5.1cm]{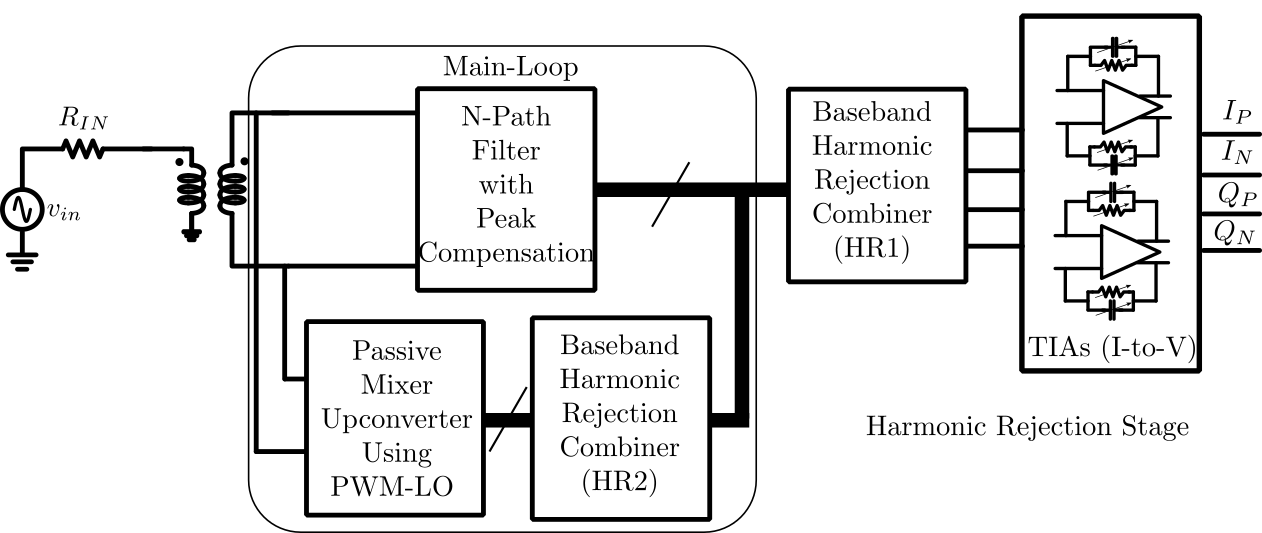}

\caption{Proposed N-path receiver with harmonic suppression \label{fig:Proposed architecture}}
}
\end{figure*}

In this work, an eight-phase, N-path receiver that suppresses the
response around the $3^{\text{rd}}$ and the $5^{\text{\text{th}}}$
harmonics of the fundamental LO frequency, $f_{LO}$, at the RF input
is proposed. The design employs capacitive baseband loads. The N-path
filter by itself provides a bandpass response at the fundamental LO
frequency and all odd harmonics of the LO, with bandwidth determined
by the baseband capacitors and the source resistance \cite{Ghaffari2010}.

\begin{figure*}[!t]
\centering{\includegraphics[width=1.8\columnwidth]{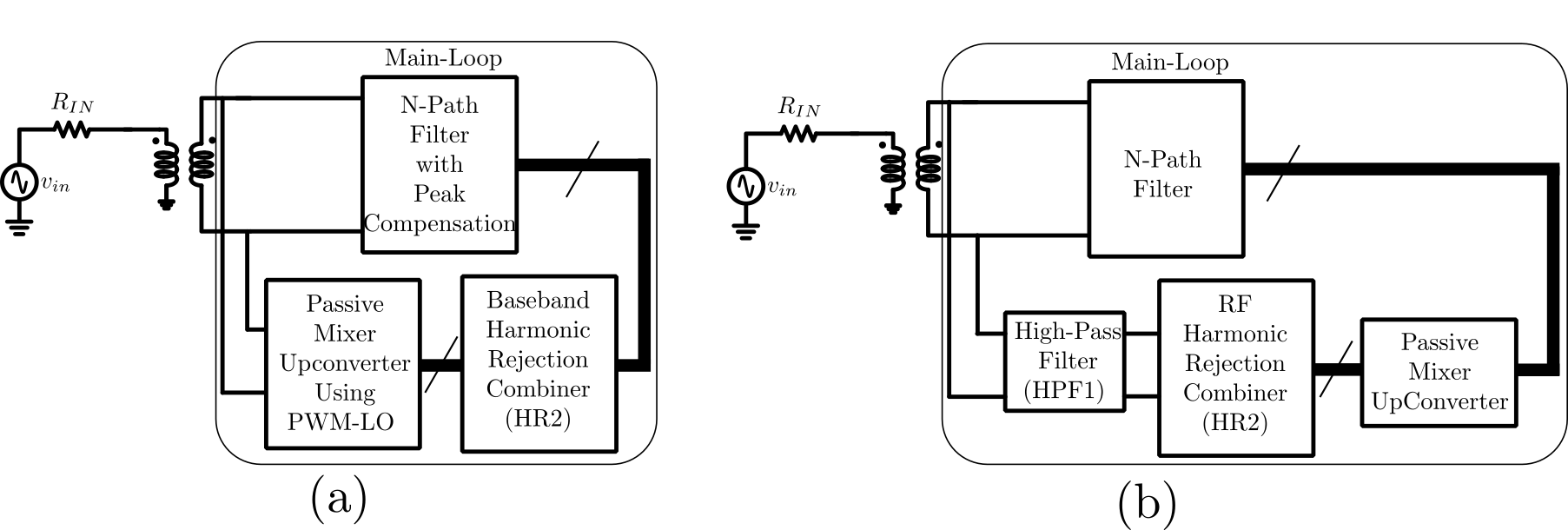}

\caption{Main loop architectures \label{fig:Main-Loop-Comparison}}
}
\end{figure*}

In order to suppress the input response from the harmonics of the
LO, the baseband signal components downconverted by the $3^{\text{rd}}$
and the $5^{\text{\text{th}}}$ harmonics of the LO signal at the
output of the N-path filter, are upconverted in a harmonic-selective
feedback path to the input. The feedback path operates only around
the $3^{\text{rd}}$ and the $5^{\text{\text{th}}}$ harmonics. This
is ensured by using a baseband harmonic selective combiner, that selects
only these signal components, and rejects the signal downconverted
by the fundamental $f_{LO}$. In addition, the feedback path upconverter
employs a pulse-width-modulated LO (PWM-LO) with an ideally zero component
around $f_{LO}$ and primary harmonics at $3f_{LO}$ and $5f_{LO}$
\cite{KangPWMHRM}\cite{KimCICC} as described below.

Suppressing the harmonic response at the RF input also ensures similar
suppression at the baseband output. Combined with another baseband
harmonic rejection stage which selects baseband signals downconverted
from the fundamental, the architecture is shown to provide harmonic
rejection of nearly a two-stage harmonic rejection downconverter \cite{ru2009HRM}\cite{forbes_design_2013}.
The filter response at $8kf_{LO}\pm f_{LO}$ is not mitigated by the
architecture, for $k\geq1$, however it assumed that the inherent
attenuation provided by the N-path filter at baseband is adequate.

The architecture is described in Section II. An approach for avoiding
the frequency-shift in the response caused by a parasitic input capacitance
\cite{pavan_npath_source_cap_2018} is also described. Section III
describes simulation-based results and the conclusion follows in Section
IV.

\section{Overview of the Architecture\label{sec:Overview-of-Architecture}}

The receiver architecture is shown in Fig. \ref{fig:Proposed architecture}.
The architecture consists of a main-loop, which provides the N-path
filtering at the input, and the baseband signals, and two harmonic
rejection combiners, one within the loop (HR2), and one outside the
loop (HR1). HR1 selects signals downconverted around $f_{LO}$, while
HR2 selects signals dowconverted from around $3f_{LO}$ and $5f_{LO}.$
The design of these is described below.

\subsection{\label{subsec:Main-Loop}The Main Loop}

The main-loop in the proposed architecture employs an N-path filter,
a baseband harmonic rejection stage (HR2) to reject the fundamental
and amplify the $3^{\text{rd}}$ and the $5^{\text{\text{th}}}$ harmonics,
and a passive-mixer based upconverter as shown in Fig. \ref{fig:Proposed architecture}.
The N-path filter consists of NMOS switches with capacitive loads
at the output, while HR2 is a transconductance-based harmonic-selection
stage, which employs $g_{m}$-ratios to suppress or enhance specific
harmonics.

HR2 suppresses the baseband signal that is downconverted from $f_{LO}$
by the N-path filter, but since it uses active devices, baseband noise,
including flicker noise, appears at its output. It is critical to
ensure that this noise does not get translated to around $f_{LO}$
at the input of the N-path filter, since this can severely degrade
the noise figure of the desired signal at this frequency. To ensure
this, the upconverter employs a pulse-width-modulated (PWM-LO) that
has an attenuated fundamental LO component, but large and nominally
equal LO components at $3f_{LO}$ and $5f_{LO}$. This clock upconverts
the baseband signal components downconverted by the N-path filter
by these harmonics, while suppressing the upconversion of baseband
noise at the output of HR2 to $f_{LO}$. Noise components around $2f_{LO}$
and $4f_{LO}$ at the output of HR2, which can also be translated
to $f_{LO}$ by the PWM-LO are suppressed by capacitance at the output
of HR2. The negative feedback provided by the loop ensures that a
low input impedance is achieved around the $3^{\text{rd}}$ and the
$5^{\text{\text{th}}}$ harmonics, thereby suppressing the signal
at this node. The differential nature of the implementation inherently
rejects even harmonics.

While a PWM-LO approach is employed here, it is also possible to upconvert
the baseband signal components from the N-path filter output to RF
using a passive mixer, and then employ harmonic selection in HR2 to
attenuate the upconverted component around $f_{LO}.$ Doing so avoids
the baseband flicker noise of HR2 from degrading the noise figure
at $f_{LO}$. However HR2 noise at $f_{LO}$ can directly appear at
the input to the N-path filter. To mitigate this, an HPF is required
in the signal path with a stop-band around $f_{LO}$ and a pass-band
around $3f_{LO}$ and $5f_{LO}$ (Fig. \ref{fig:Main-Loop-Comparison}(b)).
This filter needs to be passive as the use of active devices would
contribute additional device noise at $f_{LO}$ in the filter itself.
While an approach such as \cite{manetti_passive_1984} can be employed,
design of this filter is challenging since the loop presents a low-impedance
at the filter output at $3f_{LO}$ and $5f_{LO}$, and a larger impedance
at $f_{LO}$, which works against achieving a high-pass response from
the filter.

A more subtle issue arises if a passive mixer is directly connected
between the N-path baseband capacitors, and the inputs of the HR2
devices. The passive mixer cyclically connects the baseband outputs
of the N-path filter to any one input of the HR2 devices. This leads
to charge sharing between the baseband capacitors of the N-path filter
which should be ideally isolated from each other. This in turn leads
to a frequency shift of the input response of the N-path. This effect
is similar to the frequency shift caused by a parasitic capacitor
at the input of the N-path filter itself, as described in \cite{pavan_npath_source_cap_2018}.

For the above reasons, the approach of Fig. \ref{fig:Main-Loop-Comparison}(a)
is employed in the proposed design.

\subsubsection{\label{subsubsec:N-Path-Filter-w-PC}N-Path Filter Design}

\begin{figure}[h]
\centering\includegraphics[width=0.95\columnwidth]{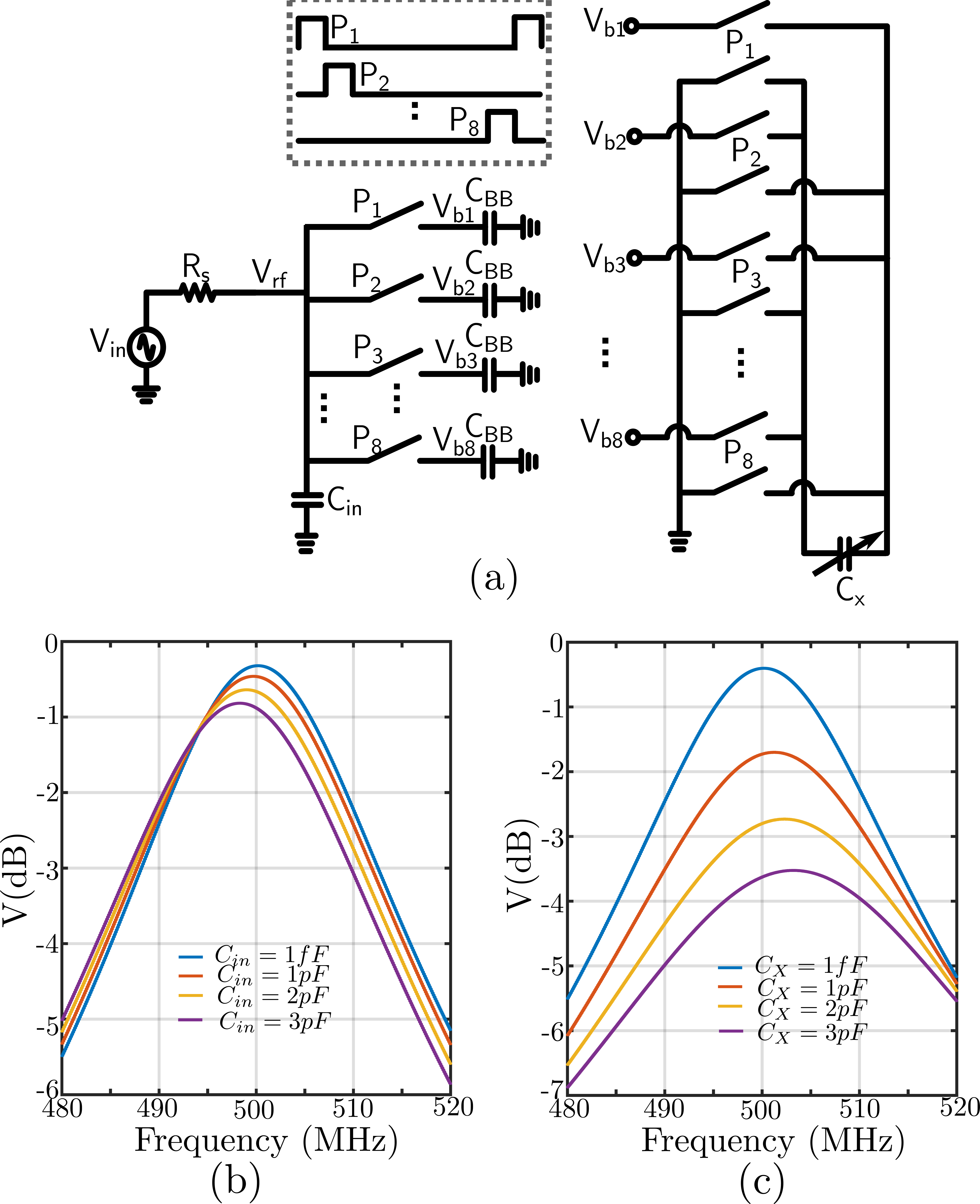}\caption{(a) N-path filter with peak compensation\label{fig:N-Path_w_PC} (b)
Frequency response of voltage $V_{rf}$ for a 0 dBV peak input $V_{in}$
with $C_{X}$ = 0 (c) Frequency response of voltage $V_{rf}$ for
a 0 dBV peak input $V_{in}$ with $C_{in}$ = 0}
\end{figure}

The main-loop in the proposed architecture employs an N-path filter
with eight-phase rectangular, non-overlapping clocks. The switches
are loaded by capacitors $C_{BB}$ and the baseband outputs $V_{b1}-V_{b8}$
are observed across these capacitors (Fig. \ref{fig:N-Path_w_PC}(a)).

It has been observed that the fundamental response at the $V_{rf}$
node in Fig. \ref{fig:N-Path_w_PC}(a) shifts left due to the parasitic
capacitance $C_{in}$ at the $V_{rf}$ node \cite{pavan_npath_source_cap_2018}
as shown in Fig. \ref{fig:N-Path_w_PC}(b), for different values of
$C_{in}.$ This frequency-shift appears because $C_{in}$ correlates
the charge held on baseband capacitors across phases $P_{1}-P_{8}.$
To compensate for the left shift of the peak due to $C_{in}$, the
circuit shown in Fig. \ref{fig:N-Path_w_PC}(a) is proposed. A toggling
capacitor $C_{X}$ is applied to the outputs $V_{b1}-V_{b8}$. $C_{X}$
is connected across $C_{BB}$, with alternating polarity as the clock
progresses from $P_{1}$ to $P_{8}$. The capacitor resets the residual
charge held by $C_{in}$ at the start of each pulse phase, thus preventing
frequency shift.\footnote{This correction is not fundamental to the functioning of the receiver
loop itself, but is beneficial for centering the frequency response.}

As shown in Fig. \ref{fig:N-Path_w_PC}(c), as $C_{X}$ increases
for $C_{in}$ = 0, the fundamental response at $V_{rf}$ shifts to
the right of $f_{LO}$. For a non-zero $C_{in}$, the frequency response
of the N-path can be centered accurately around $f_{LO}$, by choosing
appropriate value of $C_{X}$. The use of $C_{X}$ also provides loss
at the input, which is used to achieve input matching instead of an
input resistance.

A similar approach as the above, which is used to compensate the impact
of $C_{in},$ can also be employed to compensate the impact of the
parasitic capacitance at the output of the passive mixer in the approach
of Fig. \ref{fig:Main-Loop-Comparison}(b). However, a very large
value of $C_{X}$ may be required in that case, which leads to excessive
attenuation at the input of the N-path filter.

\subsubsection{\label{subsubsec:Fundamental-Suppressor}Baseband Harmonic Rejection
Combiner (HR2)}

The outputs of the N-path downconverter at baseband are combined using
the harmonic rejection combiner (HR2), which is designed to reject
the fundamental while amplifying the $3^{\text{rd}}$ and the $5^{\text{\text{th}}}$
harmonics. The harmonic rejection is performed before upconverting
the 8 baseband signals, using baseband $g_{m}$ cells with ideal gain
ratios of $-1:\sqrt{2}:-1$, implemented in practice using integer
ratios of $5:7:5$. The output of HR2 is fed to the upconverter, the
output of which is fed back to the RF input of the N-path filter.
Since the upconverted signal does not consist of a component around
the LO fundamental, the feedback-loop is ineffective at this frequency,
and has an ideal loop-gain of zero. On the other hand, the gain of
HR2 at the $3^{\text{rd}}$ and the $5^{\text{\text{th}}}$ harmonics
ensures an open-loop gain given by the product of the downconversion
and upconversion paths.

\begin{figure*}[!t]
\centering\includegraphics[height=5.9cm]{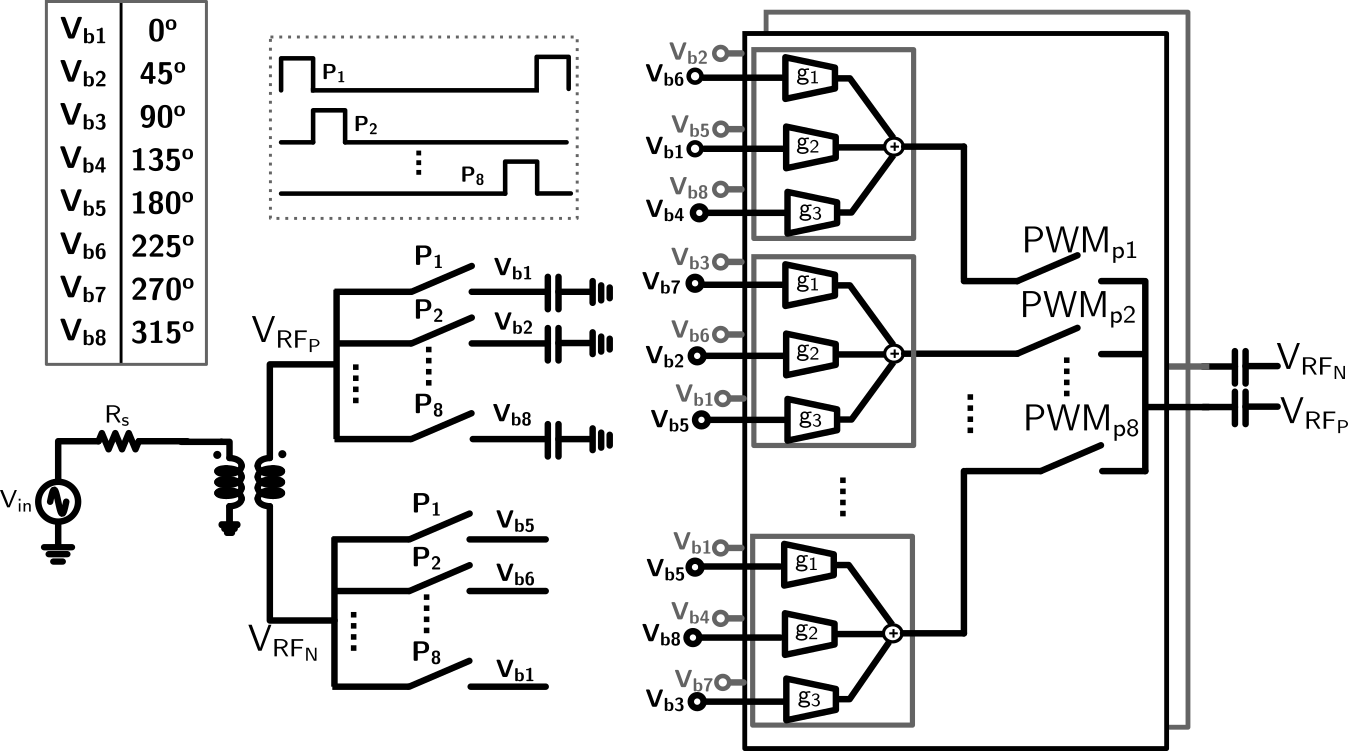}\caption{HR2: Harmonic Rejection Combiner and PWM-LO upconverter to suppress
fundamental in the feedback path\label{fig:HRM2}}
\end{figure*}

Figure \ref{fig:HRM2} represents the circuit used for implementing
the harmonic rejection mixer. The inputs to gain coefficients $g_{1}$
and $g_{3}$ are rotated by $\pm135^{o}$ compared to the inputs of
$g_{2}$ to suppress the fundamental harmonic.

\subsubsection{\label{subsubsec:Noise-Canceling}Noise Suppression Using PWM Upconverter}

As noted above, while HR2 rejects the signal component at $f_{LO}$,
it introduces baseband flicker noise at its output due to the use
of active devices in the design. The response of the upconverter around
$f_{LO}$ is suppressed by using a PWM-LO clock.

\begin{figure}
\centering\includegraphics[width=0.7\columnwidth]{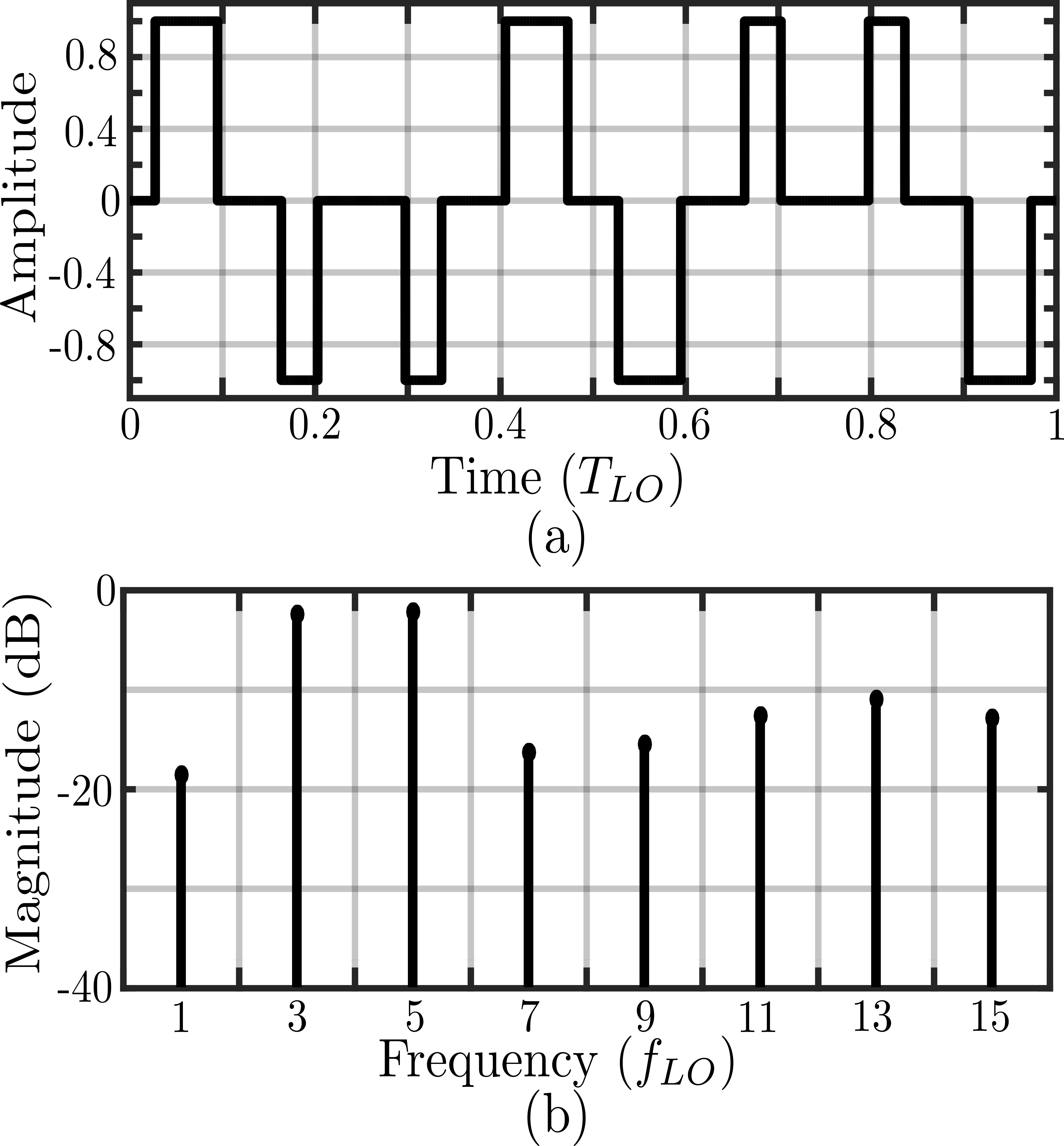}

\caption{PWM - LO (a) PWM waveform (b) Spectrum\label{fig:pwm-clock}}
\end{figure}

The PWM-LO waveform in the time-domain and its frequency response
is shown in Fig. \ref{fig:pwm-clock}. The PWM-LO is divided into
two pulse sequences, corresponding to the positive and negative polarity
of the time-domain waveform. The PWM upconverter consists of two switch
banks driven by either the positive or negative pulse sequence. For
example, $PWM_{p1}$ in Fig. \ref{fig:HRM2} represents all the positive
pulses in the PWM waveform shown in Fig. \ref{fig:pwm-clock}(a).
$PWM_{p2}$ (Fig. \ref{fig:HRM2}) is synthesized by shifting $PWM_{p1}$
by one period of the primary clock ($1/(8f_{LO})$). All the mixer
outputs clocked by $PWM_{p1},PWM_{p2},\dots,PWM_{p8}$ are combined
and applied to the positive input of the N-path. Similarly, $PWM_{n1}$
in Fig. \ref{fig:HRM2} represents all the negative pulses in the
PWM waveform shown in Fig. \ref{fig:pwm-clock}(a). $PWM_{n2}$ is
synthesized by shifting $PWM_{n1}$ by one period of the primary clock
($1/(8f_{LO}$)). All the mixer outputs clocked by $PWM_{n1},PWM_{n2},\dots,PWM_{n8}$
are combined and applied to the negative input of N-path.

An approach for generating high-frequency PWM clocks is described
in \cite{KangPWMHRM}. The approach uses a voltage-controlled delay
line (VCDL) within a delay locked loop (DLL), and can be employed
here.

\subsection{\label{subsec:HarmonicRejection}Baseband Harmonic Rejection Combiner}

A harmonic rejection combiner (HR1) is used to combine the 8-phase
baseband signals and provide differential quadrature outputs $\{I_{P}$,$I_{N}\}$
and $\{Q_{P}$,$Q_{N}\}$. The baseband combiner uses a $1:\sqrt{2}:1$
combiner \cite{weldon2001hrm}, employing suitably scaled transconductance
($g_{m}$) cells (12:17:12 in this case). The outputs of the $g_{m}$
cells are applied to transimpedance amplifiers (TIAs), that convert
the current outputs of these $g_{m}$ cells to voltages.

The feedback loop provides rejection of the harmonic responses around
$3f_{LO}$ and $5f_{LO}$ at the RF input, and at the baseband outputs
of the N-path downconverter. The use of HR1 provides effectively a
second-stage of harmonic rejection.

\begin{figure}
\centering\includegraphics[width=0.55\columnwidth]{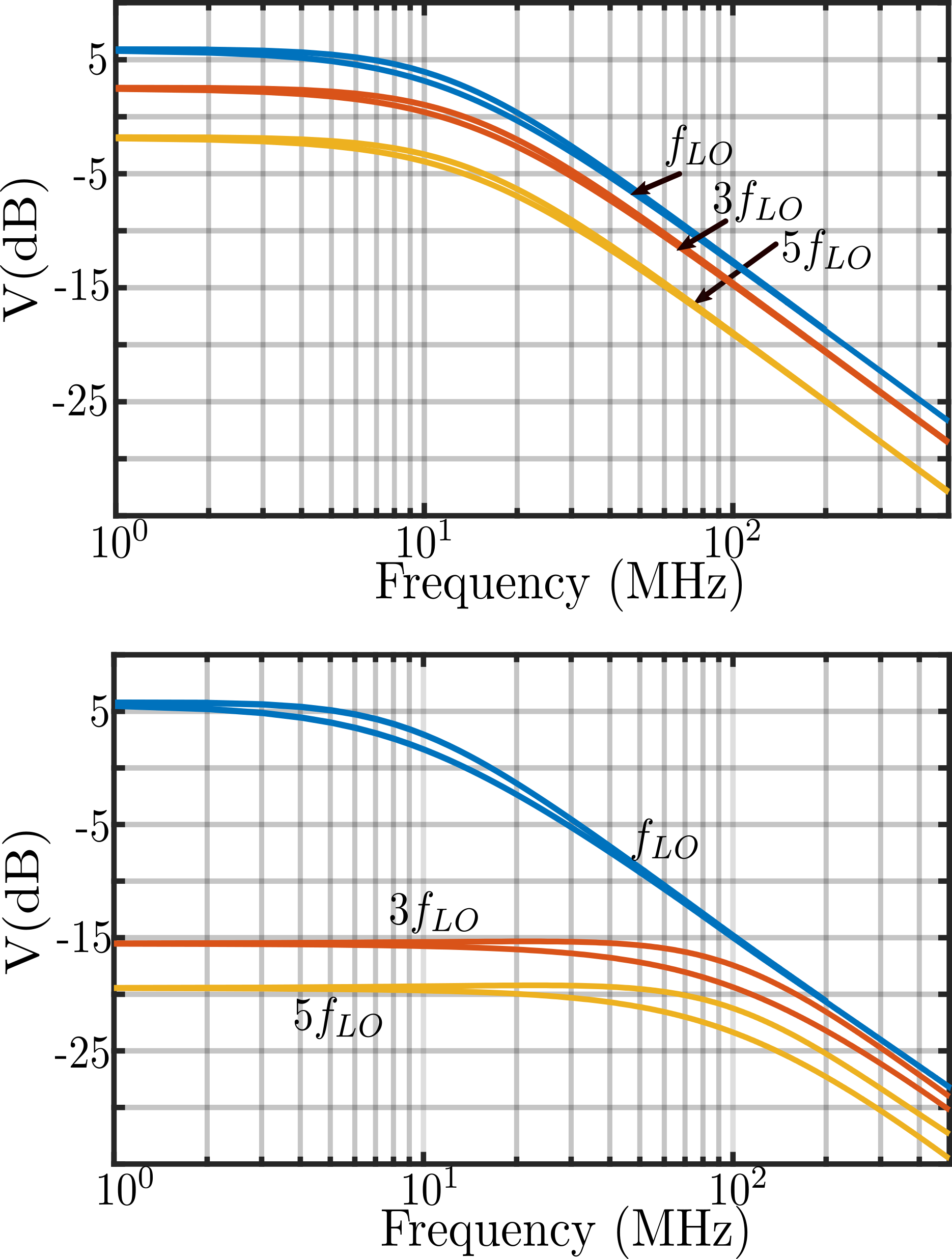}

\caption{Harmonic response at baseband for default N-path and main-loop, $f_{LO}=500$
MHz\label{fig:BB-Harmonic-response-comparison}}
\end{figure}

\section{Simulation results \label{sec:Simulation-results}}

Simulations of the proposed architecture of Fig. \ref{fig:Proposed architecture}
are shown below. The design is simulated and verified using a 65-nm
commercial CMOS process, which includes the complete device model.
The design employs an 8-phase clock with $f_{LO}$ = 500 MHz. The
harmonic responses at the baseband of an N-path filter by itself,
and for the main-loop shown in Fig. \ref{fig:Proposed architecture}
are shown in Fig. \ref{fig:BB-Harmonic-response-comparison}. The
input frequency is swept from 300 MHz - 3 GHz and the frequency-translated
output at baseband is observed. The baseband components downconverted
from $3f_{LO}$ and $5f_{LO}$ can be seen to be reduced to nearly
20 and 25 dB relative to the component from $f_{LO}$, respectively,
with the feedback loop, compared to approximately 3 dB and 6 dB respectively
for the N-path filter by itself.

\begin{figure}
\centering \includegraphics[width=0.55\columnwidth]{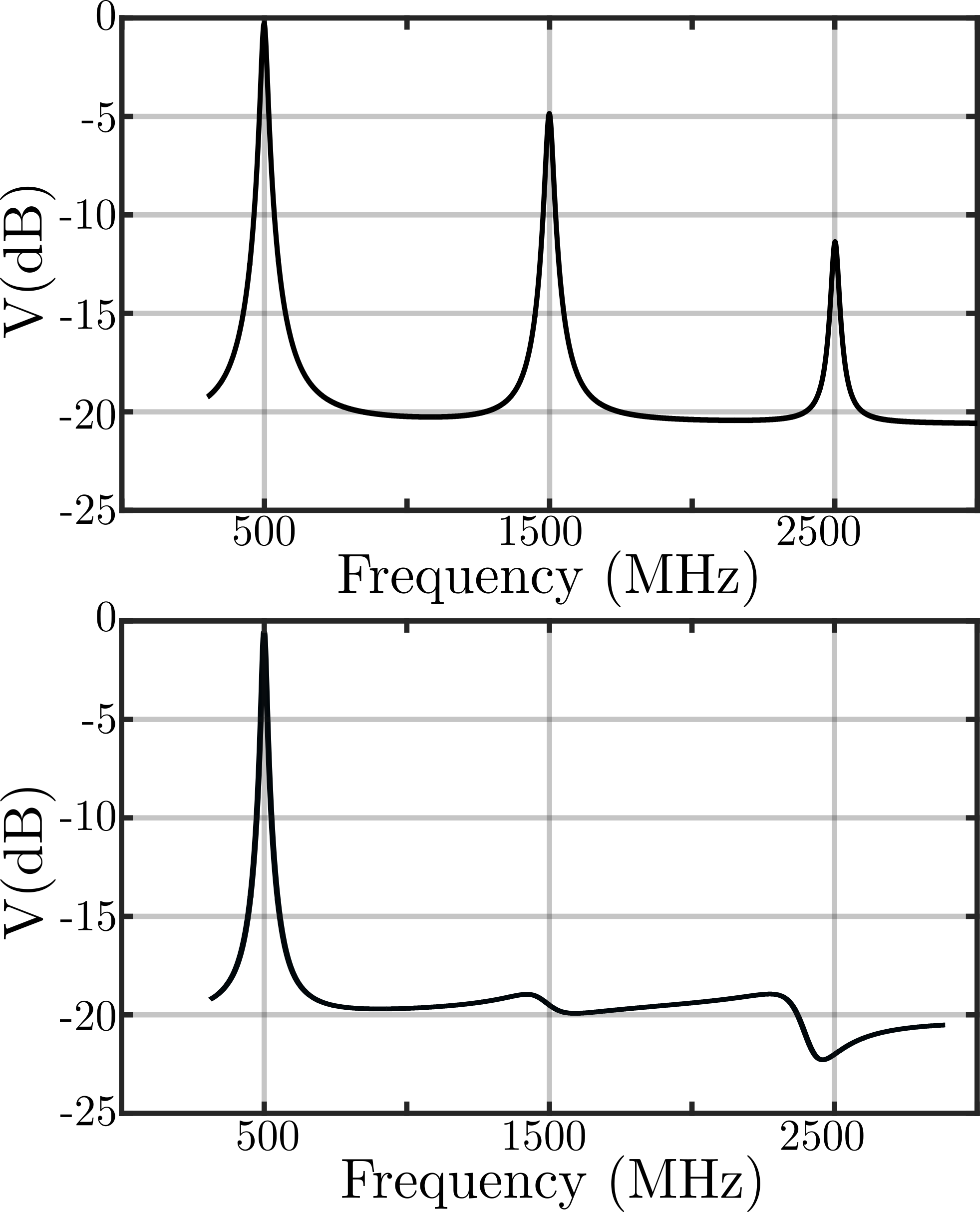}

\caption{Harmonic response at input of receiver for default N-path and main-loop,
$f_{LO}=500$ MHz\label{fig:RF-Harmonic-response-comparison}}
\end{figure}

The RF response for the above two cases is compared in Fig. \ref{fig:RF-Harmonic-response-comparison}.
The default N-path filter exhibits downconversion from all odd harmonics
of $f_{LO}$, while the main-loop shows a response only at $f_{LO}$,
and thus resembles an ideal BPF. The floor of the response is set
by the switch resistance.

\begin{figure}
\centering \includegraphics[width=0.55\columnwidth]{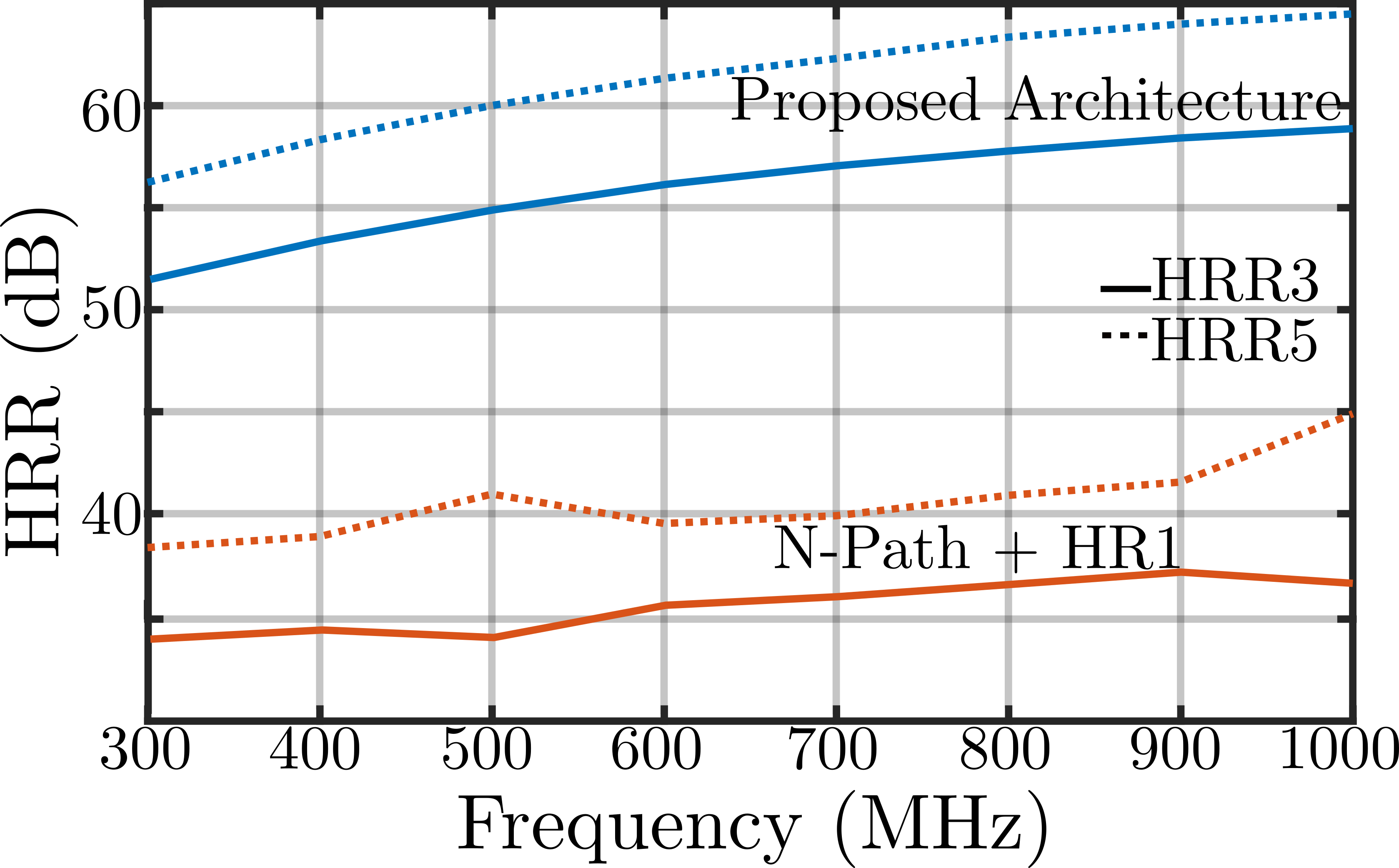}\caption{Harmonic Rejection Ratio at the $3^{\text{rd}}$ and the $5^{\text{\text{th}}}$
harmonics for proposed architecture and N-path + HR1\label{fig:HRR}}
\end{figure}

Harmonic rejection at the $3^{\text{rd}}$ and the $5^{\text{\text{th}}}$
harmonics is shown in Fig. \ref{fig:HRR}. The proposed architecture
has effectively a second-order harmonic rejection at the $3^{\text{rd}}$
and the $5^{\text{\text{th}}}$ harmonics, where as the combination
of the N-path with HR1 has only first-order harmonic rejection. The
second-order harmonic rejection is due to the suppression of the $3^{\text{rd}}$
and the $5^{\text{\text{th}}}$ harmonics at the input of the N-path
filter. A total harmonic rejection of approximately 50-60 dB is observed
in the proposed architecture. 

Simulated noise figure and S11 at $f_{LO}=500$ MHz are 6.5 dB and
-15 dB respectively. The 3-dB blocker compression point, B3dB, is
simulated by observing the blocker level around $3f_{LO}$ and $5f_{LO}$
which reduces the small-signal gain at $f_{LO}$ at the output of
HR1 by 3dB. For the proposed architecture the B3dB resulting from
the $3^{\text{rd}}$ and the $5^{\text{\text{th}}}$ harmonics of
$f_{LO}$ is +2.1 dBm and +8.4 dBm respectively. The B3dB for N-path
+ HR1, without the feedback loop, is -6.5 dBm and -3.4 dBm respectively.

\section{Conclusion \label{sec:Conclusion}}

A harmonic suppressing N-path filter that employs a a harmonic selective
combiner in the feedback path with a PWM-LO based upconverter is proposed.
A baseband harmonic combiner is employed with the feedback loop at
the output of the N-path filter, to select the desired signal from
the input spectrum. The HRM in the feedback generates the $3^{\text{rd}}$
and the $5^{\text{\text{th}}}$ harmonics responses, which effectively
attenuates the harmonic responses observed at these frequencies at
the input of the N-path filter. Due to this attenuation, the design
operates as a second-order harmonic-rejection downconverter. Simulation
results using commercial 65-nm CMOS devices are presented.

\bibliographystyle{IEEEtran}
\bibliography{arXiv_v7}

\end{document}